\documentclass[11pt]{article}

\usepackage[margin=2.5cm]{geometry}
\usepackage{amsmath,amssymb,amsthm}
\usepackage{graphicx}
\usepackage{bm}
\usepackage{hyperref}
\usepackage[utf8]{inputenc}
\usepackage[T1]{fontenc}
\usepackage{authblk}
\usepackage[version=4]{mhchem}
\usepackage[numbers]{natbib}
\title{A Net Flux-Weighted Analytical Generalization of the Greenhouse Effect Beyond the Gray Approximation}

\author{Carmine Serio}
\affil[1]{Dipartimento di Ingegneria, Università degli Studi della Basilicata, Potenza, Italy}
\affil[*]{carmine.serio@unibas.it}
\date{September 2026}

\newcommand{\Xeff}{X_{\mathrm{eff}}}

\begin{document}
	\maketitle

\begin{abstract}
	Analytical greenhouse models traditionally rely on the gray approximation, simplifying radiative transfer but treating optical depth as an unconstrained parameter that yields unrealistic surface temperatures. We derive an exact, closed-form solution directly from Schwarzschild's equations for a non-gray atmosphere. While the gray approach is physically inadequate because it over-weights saturated bands that carry negligible net energy transport, we show that non-gray formulation constrains column optical depth via the surface net radiative flux, naturally accounting for the atmospheric window without ad hoc corrections. The resulting effective optical depth penalizes saturated cores due to the negative spectral covariance between opacity and transmission. This shifts the computed surface equilibrium to an observationally consistent 288\,K, yielding a purely radiative Transient Climate Response of \(+2.18\)\,K under doubled \ce{CO2}, fully consistent with \textit{Intergovernmental Panel on Climate Change} estimates. The framework eliminates any theoretical basis for greenhouse-effect saturation. While not intended to replace three-dimensional general circulation models, this work reconciles classical radiative transfer with modern climate estimates, demonstrating that proper flux-weighting aligns analytical theory with anthropogenic warming.
\end{abstract}
	
	\section{Introduction}

The conceptual framework of planetary greenhouse warming was historically initiated by the pioneering works on radiative equilibrium in stellar atmospheres by Schwarzschild \cite{schwarzschild1906} and subsequently formalized through the linear grey-atmosphere approximations developed by Eddington \cite{eddington1916} and Milne \citep{milne1922}. When applied to terrestrial climatology, as extensively compiled in foundational texts like Houghton \citep{Houghton2002} and Goody and Yung \citep{Goody1989}, the highly structured and discontinuous infrared absorption spectrum of greenhouse gases is often replaced by this gray--atmosphere idealization. In this framework, the optical depth $\tau_{\lambda}$ (or its column-integrated counterpart) is assumed constant across all wavelengths, meaning the infrared optical thickness is represented by a single parameter that coincides with the simple arithmetic mean because the absorption coefficient is taken to be wavelength-independent. Values around two are widely used in these idealized models, following seminal studies \citep{Hopf1934, Wildt1966}.

Under a purely static radiative equilibrium, setting this unconstrained gray optical thickness to the standard textbook reference value of $\tau \approx 2$  (e.g., \cite{Houghton2002,Sagan1969}) produces a severe physical artifact, yielding an unrealistic and hydrostatically unstable surface temperature near $302$--$303\,\mathrm{K}$ ($\sim 30^{\circ}\mathrm{C}$) with an over-steep vertical lapse rate. To resolve this contradiction between radiative statics and fluid dynamics, traditional atmospheric physics has historically relied on two distinct paths. On one hand, comprehensive numerical approaches utilize radiative-convective models (RCMs), pioneered by Manabe and Strickler \citep{manabe1964}, where the reference opacity is understood as an effective, calibrated parameter that implicitly absorbs the thermodynamic stabilizing effects of latent and sensible heat fluxes \citep{hartmann2015}. On the other hand, alternative analytical formulations have attempted to modify the gray framework by introducing explicit corrections for the non-gray nature of the atmosphere; notably, Weaver and Ramanathan \citep{Weaver1995} proposed structural adjustments to account for the presence of the semi-transparent atmospheric window. Other semi-empirical approaches have instead chosen to weight the optical depth by the outgoing longwave flux at the surface \citep{Miskolczi2007, Miskolczi2010}.

Nevertheless, the discrepancy between simple radiative gray models and observed conditions has also motivated fringe frameworks that attempt to reconcile satellite observations with a saturated or self-stabilizing greenhouse effect, sometimes by imposing rigid geometric constraints on the ratio of outgoing longwave radiation (OLR) to absorbed solar flux. Such approaches typically rely on unweighted spectral averages of optical depth and can misinterpret broadband satellite data (e.g., CERES) as evidence of a locked infrared opacity.

The central aim of this paper is to show that these saturation claims stem from a fundamental mathematical misstep: relying on simple arithmetic or unweighted spectral averages of optical depth rather than a net-flux-weighted mean that reflects the actual energy transport across each spectral band. By deriving a closed-form, spectrally integrated surface-temperature relation directly from Schwarzschild's differential equations—rigorously weighting column optical depth by the net monochromatic flux—we obtain an exact non-gray generalization of the classical Milne--Houghton formulation. As demonstrated in this work, this net-flux-weighted integration naturally converges to an effective optical thickness systematically lower than its unweighted or Planck-weighted counterparts. Consequently, it shifts the calculated planetary surface equilibrium down to an observationally consistent temperate baseline ($\sim 288\,\mathrm{K}$) that is inherently coherent with anthropogenic global warming.

The paper is organized as follows. Section~\ref{section2} introduces the flux formulation of Schwarzschild’s equations and derives the non-gray greenhouse relation. Section~\ref{section3} constructs a synthetic, conceptual,  model for atmospheric opacity under hydrostatic equilibrium and uses it to demonstrate that the improved formulation yields balance parameters consistent with IPCC estimates. Section~\ref{section4} discusses the physical boundary conditions, and conclusions are drawn in Section~\ref{section5}.

\section{From Schwarzschild's differential equations to flux formulations}
\label{section2}

The derivation used in this section follows the classical textbook formulation of the greenhouse effect based on the assumption of radiative equilibrium. It is reproduced here from the beginning for the sake of clarity and to better highlight the innovative part of our approach, which is presented in Section~\ref{section2.2}.
 
Let us consider monochromatic radiative transfer through an
infinitesimal atmospheric layer of thickness $\mathrm{d}z$ and density $\rho$. Let
$I_{\lambda}(\mu,z)$ denote the spectral radiance at wavelength $\lambda$ along a direction with
cosine $\mu=\cos\theta$ relative to the vertical. Under pure absorption, the full Schwarzschild
equation including thermal emission reads
\begin{equation}
	\mu \frac{\mathrm{d} I_{\lambda}(\mu,\tau_{\lambda})}{\mathrm{d} \tau_{\lambda}}
	= - I_{\lambda}(\mu,\tau_{\lambda}) + B_{\lambda}\bigl(T(\tau_{\lambda})\bigr),
	\label{eq:schwarzschild}
\end{equation}
where $\mathrm{d}\tau_{\lambda}=\rho k_{\lambda}\mathrm{d}z$. Equation~\eqref{eq:schwarzschild}
describes directional radiance. To evaluate the planetary energy balance at a given altitude, we
transition from radiance to hemispheric fluxes.

\noindent The upward and downward monochromatic fluxes are
\begin{align}
	F_{\lambda}^{\uparrow}(\tau_{\lambda})
	&= 2\pi \int_{0}^{1} I_{\lambda}(\mu,\tau_{\lambda})\,\mu\,\mathrm{d}\mu,
	\label{eq:Fup_def} \\
	F_{\lambda}^{\downarrow}(\tau_{\lambda})
	&= 2\pi \int_{0}^{1} I_{\lambda}(-\mu,\tau_{\lambda})\,\mu\,\mathrm{d}\mu.
	\label{eq:Fdown_def}
\end{align}
The net monochromatic flux is
\begin{equation}
	\phi_{\lambda}(\tau_{\lambda}) =
	F_{\lambda}^{\uparrow}(\tau_{\lambda}) - F_{\lambda}^{\downarrow}(\tau_{\lambda}).
	\label{eq:Fnet_def}
\end{equation}

\noindent Direct substitution of Eqs.~\eqref{eq:Fup_def}–\eqref{eq:Fdown_def} into
Eq.~\eqref{eq:schwarzschild} yields a complex integro-differential system. The Elsasser
approximation simplifies this by replacing $\tau_{\lambda}$ with the flux optical depth

\[
\chi_{\lambda} = \beta\,\tau_{\lambda}, \qquad \beta = \frac{5}{3},
\]

\noindent which accounts for the average slant-path geometry of hemispheric infrared radiation. Under this
substitution, the upward and downward fluxes satisfy
\begin{align}
	\frac{\mathrm{d} F_{\lambda}^{\uparrow}}{\mathrm{d}\chi_{\lambda}}
	&= F_{\lambda}^{\uparrow} - \pi B_{\lambda}(T),
	\label{eq:Fup_diff} \\
	\frac{\mathrm{d} F_{\lambda}^{\downarrow}}{\mathrm{d}\chi_{\lambda}}
	&= - F_{\lambda}^{\downarrow} + \pi B_{\lambda}(T).
	\label{eq:Fdown_diff}
\end{align}

\noindent Optical depth is measured top-down:
\begin{equation}
	\tau_{\lambda}(z) = - \int_{+\infty}^{z} k_{\lambda}\rho\,\mathrm{d}z',
	\label{eq:optical_depth}
\end{equation}
so that it vanishes at the top of the atmosphere and reaches its maximum at the surface.

Define  the total
flux $F_{\lambda}^{\mathrm{tot}}=F_{\lambda}^{\uparrow}+F_{\lambda}^{\downarrow}$. Then, adding and
subtracting Eqs.~\eqref{eq:Fup_diff}–\eqref{eq:Fdown_diff} yields
\begin{align}
	\frac{\mathrm{d} F_{\lambda}^{\mathrm{tot}}}{\mathrm{d}\chi_{\lambda}}
	&= \phi_{\lambda},
	\label{eq:dif} \\
	\frac{\mathrm{d} \phi_{\lambda}}{\mathrm{d}\chi_{\lambda}}
	&= F_{\lambda}^{\mathrm{tot}} - 2\pi B_{\lambda}(T).
	\label{eq:sum}
\end{align}

\noindent In strict radiative equilibrium,
\begin{equation}
	\frac{\partial \phi_{\lambda}}{\partial \chi_{\lambda}} = 0,
	\label{eq:rad_eq}
\end{equation}
so Eq.~\eqref{eq:sum} implies
\begin{equation}
	F_{\lambda}^{\mathrm{tot}} = 2\pi B_{\lambda}(T).
	\label{eq:Ftot_B}
\end{equation}
Substituting into Eq.~\eqref{eq:dif} gives
\begin{equation}
	\frac{\mathrm{d}B_{\lambda}}{\mathrm{d}\chi_{\lambda}} = \frac{\phi_{\lambda}}{2\pi}.
	\label{eq:diff_B}
\end{equation}
Integrating,
\begin{equation}
	B_{\lambda} = \frac{\phi_{\lambda}}{2\pi}\chi_{\lambda} + C.
	\label{eq:diff_Bsol}
\end{equation}
The top-of-atmosphere boundary condition $\chi_{\lambda}(\infty)=0$ and
$F_{\lambda}^{\downarrow}=0$ yields $C=\phi_{\lambda}/2\pi$, giving
\begin{equation}
	B_{\lambda} = \frac{\phi_{\lambda}}{2\pi}(\chi_{\lambda} + 1).
	\label{eq:diff_Bsol1}
\end{equation}

\noindent At the surface,
\begin{equation}
	F_{\lambda}^{\uparrow}(0) = \pi B_{\lambda}(T_{g}), \qquad
	F_{\lambda}^{\mathrm{tot}}(0) = 2\pi B_{\lambda}(0),
	\label{eq:boundary}
\end{equation}
\noindent which leads to the Milne–Houghton spectral relation
\begin{equation}
	\pi B_{\lambda}(T_{g}) = \phi_{\lambda}(0)\,\bigl(0.5\,X_{\lambda} + 1\bigr),
	\label{eq:milne_spectral}
\end{equation}
\noindent where
\begin{equation}
	X_{\lambda} = -\beta \int_{\infty}^{0} \rho(z)\,k_{\lambda}(z)\,\mathrm{d}z.
	\label{eq:Xnu_def}
\end{equation}

\subsection{The gray atmosphere approximation}

Equation~\eqref{eq:milne_spectral} is spectrally resolved. Integrating over the infrared domain
$\mathcal{A}$ gives
\begin{equation}
	\int_{\mathcal{A}} \pi B_{\lambda}(T_{g})\,\mathrm{d}\lambda
	= \int_{\mathcal{A}} \phi_{\lambda}(0)\,\bigl(0.5\,X_{\lambda} + 1\bigr)\,\mathrm{d}\lambda.
	\label{eq:spectral_int}
\end{equation}
The left-hand side integrates via the Stefan–Boltzmann law,
\begin{equation}
	\int_{\mathcal{A}} \pi B_{\lambda}(T_{g})\,\mathrm{d}\lambda = \sigma T_{g}^{4}.
	\label{eq:stefan}
\end{equation}

In the gray approximation, $X_{\lambda}=X^{*}$ is taken constant. Defining
\begin{equation}
	F = \int_{\mathcal{A}} \phi_{\lambda}(0)\,\mathrm{d}\lambda,
	\label{eq:F_total}
\end{equation}
Eq.~\eqref{eq:spectral_int} becomes
\begin{equation}
	\sigma T_{g}^{4} = F\bigl(0.5 X^{*} + 1\bigr).
	\label{eq:spectral_int2}
\end{equation}

For a bare planet ($X^{*}=0$),
\begin{equation}
	F = S = \frac{S_{0}(1 - a)}{4},
	\label{eq:TOA_balance}
\end{equation}
with $S_{0}=1361\,\mathrm{W\,m^{-2}}$ and $a\approx 0.30$. The emission temperature is
\begin{equation}
	T_{0} = \left(\frac{S_{0}(1 - a)}{4\sigma}\right)^{1/4} \approx 254.58\,\mathrm{K}.
	\label{eq:T0_def}
\end{equation}
Substituting $F=S=\sigma T_{0}^{4}$ yields
\begin{equation}
	T_{g} = T_{0}\,\bigl(0.5\,X^{*} + 1\bigr)^{1/4}.
	\label{eq:Tg_final}
\end{equation}

\subsection{Non-gray atmosphere: exact spectral integration and net flux-weighted mean optical depth}
\label{section2.2}

To obtain a physically constrained global optical thickness, we introduce the net-flux-weighted mean column optical depth, $X_{\mathrm{eff}}$. Applying the integral mean value theorem to the spectrally resolved Milne--Houghton boundary relation in Eq.\eqref{eq:spectral_int}, we define:
\begin{equation}
	X_{\mathrm{eff}} = \frac{\int_{A} X_{\lambda} \, \phi_{\lambda}(0) \, \mathrm{d}\lambda}{\int_{A} \phi_{\lambda}(0) \, \mathrm{d}\lambda} = \frac{1}{F} \int_{A} X_{\lambda} \, \phi_{\lambda}(0) \, \mathrm{d}\lambda,
	\label{eq:Xeff_def}
\end{equation}
where $F = \int_{A} \phi_{\lambda}(0) \, \mathrm{d}\lambda$ represents the total integrated outgoing longwave radiative flux at the surface. The rigorous application of the theorem requires that the weighting function does not change sign within the domain of integration. This mathematical prerequisite is strictly satisfied since $\phi_{\lambda}(0) \ge 0$ across the entire infrared domain $A$, physically guaranteed because the ground temperature bounds the atmospheric thermal profile ($T_g > T(z)$), ensuring that upward surface emission everywhere exceeds downward atmospheric back-radiation.

Using the mean value theorem, the integral on the right-hand-side of Eq. Eq.\eqref{eq:spectral_int} can be solved exactly without any approximation and yields 
\begin{equation}
	\int_{\mathcal{A}} \phi_{\lambda}(0)\,\bigl(0.5\,X_{\lambda} + 1\bigr)\,\mathrm{d}\lambda=0.5\Xeff F+F=F\left (0.5\Xeff+1\right).
	\label{eq:mean_theorem_applied}
\end{equation}
and, considering also the left-hand-side of Eq.\eqref{eq:spectral_int}, we obtain  
the non-gray generalized greenhouse solution:
\begin{equation}
	\sigma T^4_g = F \left( 0.5 X_{\mathrm{eff}} + 1 \right),
\end{equation}
which can be explicitly solved for the surface equilibrium temperature as:
\begin{equation}
	T_g = T_0 \left( 0.5 X_{\mathrm{eff}} + 1 \right)^{1/4},
	\label{eq:Tg_veryfinal}
\end{equation}
where $T_0 = (F/\sigma)^{1/4} \approx 254.58\,\mathrm{K}$ denotes the effective planetary emission temperature. 

Although Eqs.~(24) and (27) share an identical algebraic structure, the physical derivation of $X_{\mathrm{eff}}$ is fundamentally distinct from the unconstrained gray parameter $X^*$. To formalize its connection to classical unstructured averages, let $L = \int_{A} \mathrm{d}\lambda$ define the absolute bandwidth of the infrared domain, and let the standard arithmetic spectral mean opacity be denoted as:
\begin{equation}
	X_m = \frac{1}{L} \int_{A} X_{\lambda} \, \mathrm{d}\lambda.
\end{equation}
By invoking the statistical definition of spectral covariance over the domain $A$, the flux-weighted optical thickness can be exactly decomposed as:
\begin{equation}
	X_{\mathrm{eff}} = X_m + \frac{L}{F} \, \mathrm{Cov}(\phi_{\lambda}, X_{\lambda}).
	\label{eq:covariance_decomp}
\end{equation}

Because strong molecular absorption lines ($\chi_\lambda \gg 1$) inherently choke radiative transmission, the net monochromatic energy transport asymptotically vanishes in saturated cores ($\phi_\lambda \to 0$). Conversely, the net transiting flux peaks exclusively within highly transparent atmospheric windows where the column opacity is minimal. This physical anti-correlation implies that $\mathrm{Cov}(\phi_{\lambda}, X_{\lambda}) < 0$, mathematically proving the general inequality:
\begin{equation}
	X_{\mathrm{eff}} < X_m.
\end{equation}

The physical origin of the negative spectral covariance behavior is also qualitatively described in the classical treatments of Goody \& Yung \cite{Goody1989} and Pierrehumbert \cite{Pierrehumbert2010}, who emphasize the role of the atmospheric window in shaping the planetary energy balance. The present work provides a formal, closed-form mathematical proof of this anti-correlation, demonstrating that it naturally penalizes saturated cores and yields an effective optical depth \(X_{\mathrm{eff}} < X_m\).

Equation~\eqref{eq:covariance_decomp} demonstrates that standard unweighted gray-atmosphere models suffer from an inherent mathematical inflation bias. By neglecting this negative spectral covariance, arithmetic averages allocate disproportionate weight to saturated absorption bands that carry negligible net energy transport, artificially shielding the planetary window leakage and causing the severe surface-temperature overestimates typically observed in textbook radiative-equilibrium formulations.

\section{Results}
\label{section3}

In this section we will describe a proof-of-concept numerical verification of the fact that  $\mathrm{Cov}(\phi_{\lambda},\chi_{\lambda})$, and then we will use the non-gray solution of the green-house effect to derive climatic implications.

\subsection{Proof-of-concept numerical verification within a discretized non-gray framework}

To provide a physically consistent bridge between the analytical formulation of Section~\ref{section2} and a realistic non-gray atmosphere, we complement the closed-form expressions with a simplified, synthetic radiative-transfer model. The main intent of this section is not to introduce an operational line-by-line or radiative-convective model, but rather to construct a lightweight and controlled numerical environment to evaluate the net flux-weighted optical depth $X_{\mathrm{eff}}$ directly from Schwarzschild's equations under representative non-gray conditions.

The demonstration algorithm is detailed in Appendix~A; here we summarize its conceptual foundation. The model does not attempt to capture the complete spectroscopic complexity of the terrestrial atmosphere, but merely serves to mimic its primary broadband opacity features \cite{Goody1989, Pierrehumbert2010}: a Gaussian-broadened $\text{CO}_2$ band centered at $15\,\mu\text{m}$, a rotational $\text{H}_2\text{O}$ continuum for $\lambda > 20\,\mu\text{m}$, and a semi-transparent baseline opacity ($\chi \approx 0.05$) representing the $8\text{--}12\,\mu\text{m}$ atmospheric window. The vibrational–rotational $6.3\,\mu\text{m}$ $\text{H}_2\text{O}$ band is intentionally omitted because, at terrestrial temperatures ($T\approx 288\,\text{K}$), the Planck emission decays rapidly at shorter wavelengths, rendering its contribution to the spectrally integrated greenhouse flux negligible relative to the $15\,\mu\text{m}$ and far-infrared regions \cite{Weaver1995, Pierrehumbert2010}.

The 1D synthetic code implements the general two-stream Schwarzschild equations over a fixed moist-adiabatic temperature profile $T(z)$. Because $T(z)$ is thermodynamically prescribed rather than iteratively solved via radiative-convective equilibrium, the net radiative flux $\phi(z)$ varies with altitude. However, this does not affect the diagnostic calculation of $X_{\mathrm{eff}}$ at the surface, which depends exclusively on the spectrally resolved surface net flux $\phi_{\lambda}(0)$ and the column optical depth $X_{\lambda}$. We emphasize that no additional physical significance is attached to this idealized numerical setup beyond its ability to faithfully reproduce the key spectral opacity distribution of the Earth's atmosphere.

\begin{figure}[htbp]
	\centering
	\includegraphics[width=\linewidth]{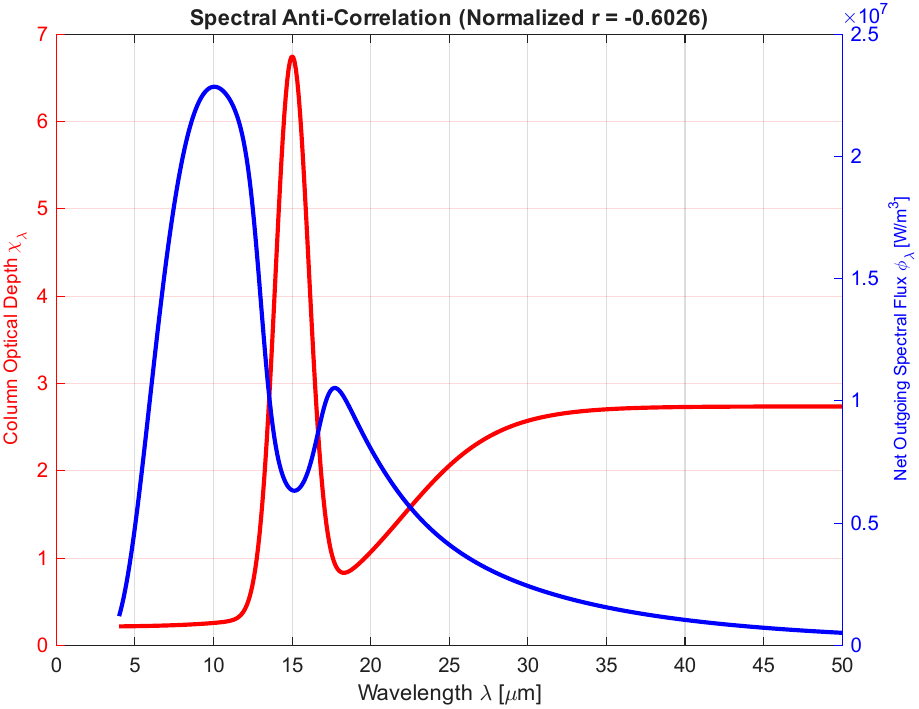}
	\caption{\textbf{Spectral anti-correlation between net outgoing flux at the surface and column optical depth.} The net spectral flux $\phi_{\lambda}$ (blue) drops significantly in regions dominated by strong greenhouse-gas absorption where $X_{\lambda}$ is large (red), yielding a strong negative correlation ($r \approx -0.60$).}
	\label{fig:spectral_anti_correlation}
\end{figure}

Figure~\ref{fig:spectral_anti_correlation} illustrates the resulting negative correlation between the outgoing net flux at the surface and the spectral optical depth: saturated absorption cores (large $X_\lambda$) inhibit net transmission ($\phi_\lambda(0) \to 0$), whereas the atmospheric window allows substantial radiative transport. The Pearson correlation coefficient, defined as
\begin{equation}
	r = \frac{\mathrm{Cov}(\phi_{\lambda},X_{\lambda})}{\sigma_{\phi_{\lambda}}\sigma_{X_{\lambda}}}
\end{equation}
yields $r = -0.67$, numerically confirming the inequality $X_{\mathrm{eff}} < X_m$. For the reference scenario considered here, $X_{\mathrm{eff}} \approx 1.27$ compared to the unweighted mean $X_m = 4.70$.

To further clarify how the scheme handles the pair $(X_{\mathrm{eff}}, X_m)$, the synthetic model contains a single global scaling parameter that can be adjusted until $T_g(X_{\mathrm{eff}}) \simeq T_s = 288\,\text{K}$, an equilibrium achieved when $X_m = 4.70$ (see Appendix~A for details).

In summary, this synthetic numerical exercise is deliberately limited in scope. It serves solely as a conceptual proof-of-concept, demonstrating that incorporating basic spectral anti-correlation into Schwarzschild's equations is sufficient to align non-gray radiative equilibrium with observed surface temperatures, fully supporting the analytical framework derived in Section~\ref{section2}.

\subsection{Climatic implications and model calibration}

Equation~\eqref{eq:Tg_veryfinal} shows that the effective global optical depth $X_{\mathrm{eff}}$ is a
dynamic variable governed by the spectral configuration of the transiting net flux. Calibrating this
analytical framework against historical and present-day climate data demonstrates its physical
robustness.

For the pre-industrial era (1850–1900), the IPCC baseline global mean surface temperature is
$T_{g,\mathrm{PI}}=286.85\,\mathrm{K}$ ($13.70^{\circ}$C). Inserting this into
Eq.~\eqref{eq:Tg_veryfinal} yields $X_{\mathrm{eff}}\approx 1.2237$.

For the present-day era (2010–2020), the observed global mean surface temperature is
$T_{g,\mathrm{PD}}=288\,\mathrm{K}$ ($14.84^{\circ}$C). The corresponding effective optical depth is
$X_{\mathrm{eff}}\approx 1.2757$, a $\sim 4.3\%$ increase relative to the pre-industrial baseline.

We now evaluate the climate response to a doubling of atmospheric \ce{CO2}. Because the central
lines of the $15\,\mu\mathrm{m}$ \ce{CO2} band are saturated, additional \ce{CO2} increases opacity
primarily through pressure broadening and expansion of the band wings. Radiative-transfer theory
dictates that this wing expansion scales logarithmically with the volume mixing ratio $C$
\cite{Romps2022}:
\begin{equation}
	X_{\mathrm{eff}}(C) = X_{\mathrm{eff},0} + \alpha\,\ln\!\left(\frac{C}{C_{0}}\right),
	\label{eq:log_chi}
\end{equation}
where $C_{0}$ and $X_{\mathrm{eff},0}$ are the baseline concentration and effective optical depth,
and $\alpha$ is the intrinsic sensitivity coefficient. Further details about the rationale of this logarithmic scaling is given in Appendix B.

Using the two empirical reference points, the observed modern opacity increment is $\Delta X_{\text{eff}} = 1.2757 - 1.2237 = 0.0520$, which isolates the intrinsic climate sensitivity coefficient as:
\begin{equation}
	\alpha = \frac{0.0520}{\ln(400/278)} \approx 0.1429.
\end{equation}
For a pre-industrial doubling of atmospheric \ce{CO2} ($C_{2\times} = 556$~ppm), the logarithmic expansion of the band wings shifts the effective column thickness to:
\begin{equation}
	X_{\text{eff}}(556\,\mathrm{ppm}) = 1.2237 + 0.1429 \ln(2) = 1.3228.
\end{equation}
Inserting this flux-weighted opacity into the non-gray generalized radiative equilibrium relation yields the unperturbed doubled-\ce{CO2} surface equilibrium state:
\begin{equation}
	T_{s,2\times} = 254.58 \left( 1 + \frac{1.3228}{2} \right)^{1/4} = 289.03\,\mathrm{K} \quad (15.88^{\circ}\mathrm{C}).
\end{equation}
Consequently, this formulation isolates a pure Transient Climate Response (TCR) of:
\begin{equation}
	\text{TCR} = T_{s,2\times} - T_{g,\text{PI}} = 289.03\,\mathrm{K} - 286.85\,\mathrm{K} = +2.18\,\mathrm{K}.
\end{equation}
This analytical result is summarized in Table~\ref{tab:climate_states} and is fully consistent with the independent multi-model central estimates compiled in the IPCC AR6 assessment.

\begin{table}[ht]
	\caption{Logarithmic effective optical-depth scaling and derived surface equilibrium temperatures.}
	\label{tab:climate_states}
	\centering
	\begin{tabular}{lccc}
		\hline
		State & \ce{CO2} Concentration & Effective Opacity $X_{\text{eff}}$ & Surface Temp. $T_s$ [K] ($^{\circ}\mathrm{C}$) \\
		\hline
		Pre-industrial   & 278 ppm & 1.2237 & 286.85 (13.70) \\
		Modern (2010--2020) & 400 ppm & 1.2757 & 288.00 (14.85) \\
		Doubled $2\times\ce{CO2}$ & 556 ppm & 1.3228 & 289.03 (15.88) \\
		\hline
	\end{tabular}
\end{table}
By relying exclusively on the logarithmic response of the flux-weighted optical depth, this
formulation shows that $X_{\mathrm{eff}}$ relaxes from its pre-industrial baseline of $1.2237$ to
$1.3166$ under $2\times\ce{CO2}$, producing a purely radiative warming of $+2.18\,\mathrm{K}$,
fully consistent with current IPCC estimates. In IPCC AR6 terminology, this instantaneous
surface-temperature increase corresponds to the Transient Climate Response (TCR), whose
very likely range is $1.2^{\circ}\mathrm{C}$ to $2.4^{\circ}\mathrm{C}$, with a central estimate of
$1.8^{\circ}\mathrm{C}$~\cite{IPCC_AR6_Ch7,IPCC2021SPM}.

It should be noted that the results shown in this section rely only on the application od the basic Eq. \eqref{eq:Tg_veryfinal}, which is independent on any numerical scheme for the calculations, e.g., of net fluxes.

\section{Discussion}
\label{section4}

The analytical relation in Eq.~\eqref{eq:Tg_veryfinal} describes a steady-state radiative
equilibrium and is therefore optimized to capture short- to medium-term climate responses
(TCR). Long-term, nonlinear slow feedbacks—such as deep-ocean heat uptake, cryospheric
changes, or large-scale ice–albedo feedbacks (ECS)—are not explicitly represented and would
require coupling to dynamical and thermodynamic models.

As emphasized earlier, the use of $\phi_{\lambda}$ as a weighting function relies on its
non-negativity across the spectrum. In practice, the spectral profile of $\phi_{\lambda}$ can be
obtained directly from high-resolution spaceborne infrared spectrometers (e.g., IASI, AIRS)
co-located with CERES broadband measurements. This observational anchoring transforms
Eq.~\eqref{eq:Tg_veryfinal} from a purely theoretical construct into a diagnostic tool capable of
mapping observed TOA spectral fluxes into integrated planetary surface temperatures.

The total column optical depth $X_{\lambda}$ is defined as an exact vertical integral from
$z=0$ to $z\rightarrow\infty$, and there is no implicit or explicit limitations to its vertical structure. 
structure of the atmosphere.

It is important to distinguish the present formulation from established mean-opacity methods. Standard gray approaches replace the non-gray problem by defining a spectrally averaged opacity, such as the Rosseland mean (appropriate for the diffusive interior) or the Planck mean (appropriate for optically thin regions). Semi-empirical non-gray approaches, in contrast, retain spectral structure but introduce explicit corrections for the atmospheric window \cite{Weaver1995}. The present formulation differs from both: it derives an effective optical depth directly from the net spectral flux via the integral mean value theorem, without invoking a mean opacity or requiring structural corrections.

The Rosseland mean optical depth is defined as (e.g., \cite{Rosseland1924})
\begin{equation}
	\frac{1}{\tau_R} = \frac{\displaystyle\int_0^\infty \frac{1}{\tau_\lambda} \frac{\partial B_\lambda}{\partial T} \, d\lambda}{\displaystyle\int_0^\infty \frac{\partial B_\lambda}{\partial T} \, d\lambda},
	\label{eq:rosseland}
\end{equation}
and is designed to yield the correct radiative flux in the diffusion limit, where the radiation field is nearly isotropic and the atmosphere is optically thick. It weights spectral regions inversely by optical depth, so that transparent windows dominate the harmonic mean.

The Planck mean optical depth is instead defined as (e.g., \cite{Planck1901})
\begin{equation}
	\tau_P = \frac{\displaystyle\int_0^\infty \tau_\lambda B_\lambda \, d\lambda}{\displaystyle\int_0^\infty B_\lambda \, d\lambda},
	\label{eq:planck}
\end{equation}
and is appropriate for optically thin emission, where the source function is Planckian and the optical depth is weighted by the local emission spectrum. A Planck-like weighted mean is used, for example, in the work of \cite{Miskolczi2007,Miskolczi2010}, who report an observationally supported total optical depth of approximately 1.87.

Neither of these means is defined as a net flux-weighted average of the optical depth. In particular, the Rosseland mean is a harmonic mean weighted by the temperature derivative of the Planck function, while the Planck mean is an arithmetic mean weighted by the Planck function itself. Neither incorporates the net spectral flux \(\phi_\lambda(0)\) that actually transports energy through the atmospheric column. Consequently, they cannot capture the negative spectral covariance between opacity and net transmission that is central to the present derivation (Eq.~\eqref{eq:covariance_decomp}). The effective optical depth \(X_{\mathrm{eff}}\) introduced in Eq.~\eqref{eq:Xeff_def} is therefore fundamentally distinct from both classical means: it is a direct, net flux-weighted average of the spectral optical depth, derived rigorously from Schwarzschild's equations and the integral mean value theorem.

The gray atmosphere approach has also been abandoned because of its failure to impose any limit on the growth of the optical depth, thereby permitting a runaway greenhouse effect in the limit \(\tau_\lambda \to +\infty\). In our non-gray approach, the infinite growth of \(\tau_\lambda\) is bounded: the net-flux-weighted average tends to zero within strongly absorbing bands, while remaining finite in atmospheric window regions.

This latter behavior might appear to invoke a saturation effect. However, such saturation is physically excluded by the fact that the most important atmospheric window, 8--12~\(\mu\)m, also contains weak \ce{CO2} absorption bands---the so-called hot bands (or, in laser physics, laser bands). At sufficiently high \ce{CO2} concentrations, these bands strengthen and progressively close the window \cite{Kratz1991,Chou1993}. Consequently, \(X_{\mathrm{eff}}\) cannot saturate: as \ce{CO2} increases, the window itself becomes partially opaque, ensuring that the flux-weighted optical depth remains responsive to further \ce{CO2} increases.

The framework developed here provides a natural diagnostic for quantifying the radiative impact of these hot bands. Because \(X_{\mathrm{eff}}\) is defined as a net-flux-weighted average of the spectral optical depth, any strengthening of the hot bands within the window directly translates into an increase in \(X_{\mathrm{eff}}\), and hence into a warming of the surface equilibrium temperature through Eq.~\eqref{eq:Tg_veryfinal}. This offers a transparent analytical route to assess the potential danger of hot-band enhancement under extreme \ce{CO2} scenarios, a regime in which the atmospheric window can no longer be treated as a fixed, transparent channel. In this sense, the present formulation not only reconciles classical radiative transfer with observed climate sensitivity but also provides a physically grounded tool for exploring the nonlinear, window-closing regimes that three-dimensional models must resolve numerically.   

\section{Conclusion}
\label{section5}

By deriving the analytical formulation directly from Schwarzschild's differential equations and employing the integral mean value theorem with the net spectral flux as a weighting function, we have established a mathematically exact, non-gray generalization of the classical Milne--Houghton radiative equilibrium model. This formulation resolves the artificial surface-temperature overestimation ($\sim 302\,\mathrm{K}$) inherent to standard unweighted gray-atmosphere treatments, which allocate disproportionate weight to saturated absorption bands that carry negligible net energy transport.

The central physical insight lies in recognizing that the intrinsic negative spectral covariance between column opacity and net radiative transmission naturally dampens the radiative weight of heavily saturated infrared cores. Consequently, the effective optical depth required to reproduce the observed global mean surface temperature under radiative equilibrium is systematically lower than traditional textbook values, which rely on the gray approximation or on unweighted and Planck-weighted spectral means.

As emphasized throughout this work, this simplified, vertically integrated framework is not intended to replace or compete with sophisticated three-dimensional General Circulation Models (GCMs) that explicitly resolve fluid dynamics, cloud microphysics, and non-radiative turbulent heat fluxes. Instead, it addresses a fundamental conceptual vulnerability in basic radiative theory by providing a mathematically closed and physically sound proof that is structurally responsive to greenhouse-gas increases. 

In doing so, this framework refutes fringe saturation hypotheses that misinterpret broadband satellite observations as evidence of a locked, static greenhouse effect. Ultimately, it demonstrates that classical textbook radiative transfer—when integrated under proper net-flux-weighted operators—is inherently self-consistent and fully aligned with the observed reality of anthropogenic global warming.

\appendix

\appendix
\section*{Appendix A: Synthetic Non-Gray Radiative-Transfer Model}

The synthetic non-gray radiative-transfer model used in Section~3.1 follows a long-established 
tradition in analytical and semi-analytical infrared radiative modeling. Classical treatments of 
non-gray absorption by CO$_2$ and H$_2$O \citep{Elsasser1942,Goody1989} 
represent the 15\,$\mu$m CO$_2$ band using smooth Gaussian envelopes and the far-infrared 
H$_2$O rotational continuum using logistic or exponential activation functions. The semi-transparent 
8--12\,$\mu$m atmospheric window is likewise treated as a residual baseline in analytical frameworks 
\citep{Weaver1995,Pierrehumbert2010}. Modern wideband LBL parametrizations 
\citep{MatsunobuCoimbra2024} adopt the same conceptual structure: a smooth CO$_2$ band, a 
continuum-like H$_2$O contribution, and a window region that dominates net transmission. 
The synthetic model below is therefore  an ad hoc construction, but capable of  
representing the dominant infrared features of Earth’s greenhouse spectrum.

\subsection*{A.1 Vertical discretization and thermodynamic structure}

The atmospheric column extends from the surface ($z=0$) to a prescribed top-of-troposphere 
height $z_{\mathrm{top}}$. It is partitioned into $N$ homogeneous layers of thickness
\begin{equation}
	\Delta z = \frac{z_{\mathrm{top}}}{N}.
	\label{eq:A1}
\end{equation}

The midpoint altitude of layer $j$ is
\begin{equation}
	z_j = \left(j-\tfrac{1}{2}\right)\Delta z, \qquad j=1,\dots,N.
		\label{eq:A2}
\end{equation}

Temperature decreases with height according to a moist-adiabatic lapse rate $\Gamma$:
\begin{equation}
	T_j = T_{\mathrm{surf}} - \Gamma\, z_j.
		\label{eq:A3}
\end{equation}

The relative air-density profile is represented by an exponential stratification with scale height $H$:
\begin{equation}
	\rho_j = \exp\!\left(-\frac{z_j}{H}\right).
		\label{eq:A4}
\end{equation}

\subsection*{A.2 Spectral discretization and synthetic extinction profile}

The infrared spectrum is discretized into $M$ channels with wavelengths $\lambda_k$ spanning 
the longwave domain:
\begin{equation}
	\lambda_k \in [\lambda_{\min},\,\lambda_{\max}], \qquad k=1,\dots,M.
\end{equation}

The synthetic extinction profile combines the dominant absorption features of CO$_2$ and H$_2$O 
together with a semi-transparent atmospheric window. The total extinction coefficient is
\begin{equation}
	\sigma_{\mathrm{tot}}(\lambda_k) 
	= A_{\mathrm{ext}}\left[
	A_{\mathrm{CO_2}}\,\exp\!\left(-\frac{(\lambda_k-\lambda_{\mathrm{CO_2}})^2}{2\,\delta_{\mathrm{CO_2}}^2}\right)
	+ \frac{A_{\mathrm{H_2O}}}{1+\exp\!\left(-\frac{\lambda_k-\lambda_{\mathrm{H_2O}}}{\delta_{\mathrm{H_2O}}}\right)}
	+ A_{\mathrm{win}}
	\right].
		\label{eq:A5}
\end{equation}

\subsection*{A.3 Layer optical depth}

The infinitesimal optical-depth increment in layer $j$ and channel $k$ is
\begin{equation}
	\Delta\tau_{j,k} = \sigma_{\mathrm{tot}}(\lambda_k)\,\rho_j\,\Delta z_{\mathrm{km}},
		\label{eq:A6}
\end{equation}
where $\Delta z_{\mathrm{km}}$ converts the layer thickness from meters to kilometers.

The column optical depth for channel $k$ is
\begin{equation}
	X_{\lambda_k} = \beta\sum_{j=1}^{N} \Delta\tau_{j,k},
	\label{eq:A7}
\end{equation}
where $\beta$ is the Elsasser diffusivity factor \citep{Elsasser1942}.

\subsection*{A.4 Two-stream Schwarzschild closure}

Let $B(\lambda_k,T_j)$ denote the monochromatic Planck radiance. The hemispheric two-stream 
source term is
\begin{equation}
	S_{j,k} = \pi\,B(\lambda_k,T_j).
		\label{eq:A8}
\end{equation}

Upward flux at interface $j+1$:
\begin{equation}
	F^{+}_{j+1,k} = F^{+}_{j,k}\,\exp(-\beta\,\Delta\tau_{j,k})
	+ S_{j,k}\left[1-\exp(-\beta\,\Delta\tau_{j,k})\right].
		\label{eq:A9}
\end{equation}

Downward flux at interface $j$:
\begin{equation}
	F^{-}_{j,k} = F^{-}_{j+1,k}\,\exp(-\beta\,\Delta\tau_{j,k})
	+ S_{j,k}\left[1-\exp(-\beta\,\Delta\tau_{j,k})\right].
		\label{eq:A10}
\end{equation}

Boundary conditions:

Top of atmosphere:
\begin{equation}
	F^{-}_{N+1,k}=0.
		\label{eq:A11}
\end{equation}

Surface:
\begin{equation}
	F^{+}_{1,k} = \pi\,B(\lambda_k,T_{\mathrm{surf}}).
		\label{eq:A12}
\end{equation}

Net spectral flux at the surface:
\begin{equation}
	\phi_{\lambda_k} = F^{+}_{1,k} - F^{-}_{1,k}.
		\label{eq:A13}
\end{equation}

\subsection*{A.5 Broadband optical depths}

Unweighted arithmetic mean:
\begin{equation}
	X_m = \frac{1}{M}\sum_{k=1}^{M} X_{\lambda_k}.
		\label{eq:A14}
\end{equation}

Flux-weighted mean:
\begin{equation}
	X_{\mathrm{eff}} = 
	\frac{\sum_{k=1}^{M} \phi_{\lambda_k}\,X_{\lambda_k}}
	{\sum_{k=1}^{M} \phi_{\lambda_k}}.
		\label{eq:A15}
\end{equation}

\subsection*{A.6 Numerical parameter values}

For the simulations presented in Section~3.1, the symbolic parameters above are instantiated as:

\begin{itemize}
	\item $z_{\mathrm{top}} = 15\,\mathrm{km}$, \quad $N = 50$ \quad $\Rightarrow$ \quad $\Delta z = 300\,\mathrm{m}$,
	\item $T_{\mathrm{surf}} = 288\,\mathrm{K}$, \quad $\Gamma = 6.5\times 10^{-3}\,\mathrm{K\,m^{-1}}$,
	\item $H = 8\,\mathrm{km}$,
	\item $\lambda_{\min} = 4\,\mu\mathrm{m}$, \quad $\lambda_{\max} = 50\,\mu\mathrm{m}$, \quad $M = 500$,
	\item $A_{\mathrm{ext}} = 0.831\,\mathrm{km^{-1}}$,
	\item $A_{\mathrm{CO_2}} = 1.5$, \quad $A_{\mathrm{H_2O}} = 0.6$, \quad $A_{\mathrm{win}} = 0.05$,
	\item $\lambda_{\mathrm{CO_2}} = 15\,\mu\mathrm{m}$, \quad $\delta_{\mathrm{CO_2}} = 1.5\,\mu\mathrm{m}$,
	\item $\lambda_{\mathrm{H_2O}} = 22\,\mu\mathrm{m}$, \quad $\delta_{\mathrm{H_2O}} = 3\,\mu\mathrm{m}$,
	\item $\beta = 5/3 \approx 1.66$.
\end{itemize}

\noindent The relative amplitudes \(A_{\mathrm{CO}_2} = 1.5\), \(A_{\mathrm{H}_2\mathrm{O}} = 0.6\), and \(A_{\mathrm{win}} = 0.05\) are not derived from first-principles spectroscopy but are chosen to reproduce the qualitative spectral hierarchy of Earth's infrared absorption: a dominant CO\(_2\) band at 15~\(\mu\)m, a significant H\(_2\)O rotational continuum beyond 22~\(\mu\)m, and a residual window absorption. This choice is consistent with the classical band-model tradition of Elsasser \cite{Elsasser1942} and Goody \& Yung \cite{Goody1989,Houghton2002,Pierrehumbert2010}, where Gaussian envelopes and logistic activation functions represent the dominant spectral features. The global amplitude \(A_{\mathrm{ext}} = 0.831\)~km\(^{-1}\) is then calibrated to ensure that $T_g(\Xeff)$ yields the benchmark surface temperature $T_{surf}=288$.  A sensitivity analysis (not shown) confirms that the qualitative result \(X_{\mathrm{eff}} < X_m\) is robust to reasonable variations in these relative amplitudes, as it depends primarily on the spectral anti-correlation between opacity and net flux, not on the precise numerical values.

\appendix
\section*{Appendix B: Rationale logarithmic scaling \ce{CO2} doubling}

To justify the logarithmic scaling rule used for the $\ce{CO2}$ doubling scenario, we describe the underlying physical rationale following the classical band-wing framework. Modern theoretical applications of such an approach have been recently utilized by Romps et al. \cite{Romps2022} to demonstrate the logarithmic scaling laws of saturated greenhouse gases. 

Consider a non-gray infrared absorption band whose wing absorption coefficient decays exponentially away from the band center:
\begin{equation}
	k(x; C) = C k_0 e^{-\alpha|x|},
\end{equation}
where $C$ is the $\ce{CO2}$ concentration, $k_0$ a reference absorption amplitude, and $\alpha > 0$ the spectral decay rate. The corresponding monochromatic optical depth is:
\begin{equation}
	\tau(x; C) = k(x; C) m = C k_0 m e^{-\alpha|x|},
\end{equation}
with $m$ representing the atmospheric column mass. We define $x_c(C)$ as the precise spectral location where the band transitions from saturated to partially transparent, corresponding to the optical boundary:
\begin{equation}
	\tau(x_c; C) = 1.
\end{equation}
Solving Eq. (55) for the coordinate $x_c(C)$ yields:
\begin{equation}
	x_c(C) = \frac{1}{\alpha} \ln(C k_0 m).
\end{equation}
Since radiative forcing and flux-weighted opacity are governed by the spectral expansion and width of this transition region, the effective optical depth responds proportionally to the shift of $x_c(C)$, which directly results in:
\begin{equation}
	\tau_{\mathrm{eff}}(C) = \tau_{\mathrm{eff},0} + a \ln\left(\frac{C}{C_0}\right),
\end{equation}
where $a$ is an empirical constant determined by the specific band-wing structure and the atmospheric column properties. Under exponentially decaying band-wing absorption, the effective optical depth depends logarithmically on the greenhouse gas concentration.

\end{document}